\documentclass[conference]{IEEEtran}
\usepackage[utf8]{inputenc}
\usepackage{amsmath,amssymb,amsfonts}
\usepackage{algorithmic}
\usepackage{graphicx}
\usepackage{textcomp}
\usepackage{xcolor}
\usepackage{url}
\usepackage{float}
\usepackage{tabularx}
\usepackage{multirow}

\usepackage{microtype}

\usepackage[numbers,square]{natbib}

\usepackage{booktabs}
\usepackage{color}
\usepackage{fancyhdr}
\usepackage{atbegshi}
\usepackage{geometry}

\fancypagestyle{firstpage}{
  
  \lhead{}
  \rhead{}
  \lfoot{}
}
\fancypagestyle{rest}{%
  \lhead{Graf, Opara and Barthet}
  \rhead{An Audio-Driven System For Real-Time Music Visualisation}
}

\title{An Audio-Driven System For Real-Time Music Visualisation}

\author{\IEEEauthorblockN{Max Graf, Harold Chijioke Opara, Mathieu Barthet}
\IEEEauthorblockA{
Centre for Digital Music, Queen Mary University of London\\
Correspondence should be addressed to Max Graf (\url{max.graf@qmul.ac.uk})\\
}
}

\begin{document}
\maketitle
\thispagestyle{firstpage}
\pagestyle{rest}

Disclaimer: this paper is an Author’s Accepted Manuscript, the published version is available at \url{https://www.aes.org/e-lib/browse.cfm?elib=21091}\\ 
Main reference: \textit{M. Graf, H. Chijioke Opara, and M. Barthet, "An Audio-Driven System for Real-Time Music Visualisation," Paper 10498, (2021 May.).}\\
\textcopyright\ CC-BY 4.0\\

\begin{abstract}
Computer-generated visualisations can accompany recorded or live music to create novel audiovisual experiences for audiences. We present a system to streamline the creation of audio-driven visualisations based on audio feature extraction and mapping interfaces. Its architecture is based on three modular software components: backend (audio plugin), frontend (3D game-like environment), and middleware (visual mapping interface). We conducted a user evaluation comprising two stages. Results from the first stage (34 participants) indicate that music visualisations generated with the system were significantly better at complementing the music than a baseline visualisation. Nine participants took part in the second stage involving interactive tasks. Overall, the system yielded a Creativity Support Index above average (68.1) and a System Usability Scale index (58.6) suggesting that ease of use can be improved. Thematic analysis revealed that participants enjoyed the system's synchronicity and expressive capabilities, but found technical problems and difficulties understanding the audio feature terminology.
\end{abstract}

\section{Introduction}
Although the primary mode of consumption of music is auditory, efforts have been made to translate musical expression to other domains \cite{olowe2016featur}. For people with normal or corrected-to-normal vision, visual representations can lead to deeper insights about musical expression.

This work is concerned with computer-generated visualisation of audio signals in a 3D visual environment. Current tools to produce music visualisations do not explicitly leverage advances made in the music information retrieval (MIR) field. We investigate methods for interactively mapping audio features (numerical data representing signal and perceptual attributes of sound/music obtained computationally) to visual objects to facilitate music visualisation. The main contributions of this work are as follows: First, we propose a method to enable users to create visualisations from audio signals routed to a digital audio workstation (DAW), and a modular software system called ``Interstell.AR'', which is versatile and flexible and can be used with many game engines. Embedded in a DAW, a source can be, for example, an instrument, a singer's voice, or a complete recording. Second, we provide design insights based on a user evaluation conducted with 34 participants that can inform the design of multimedia systems combining audio and visual modalities.

As a tool to facilitate the creation of new audiovisual media, the proposed system could be of interest to musical artists, graphic designers, and VJs (visual jockeys). Applications range from the production of live video-based music streaming, visuals for live music, augmented/virtual/mixed reality experiences, games, etc. Other applications include music education, e.g., to learn about musical attributes or a performer's musical expression using visual correlates. 

Understanding the role of music visuals has been the object of studies in performance reception and experience and new interfaces for musical expression, some of which specifically focus on electronic music \cite{correia2017role}. In live electronic music performance, often the gestures of performers do not provide explicit connections to the sound production mechanisms due to the virtualisation of musical instruments (use of synthesis and digital music interfaces); accompanying visuals may be used to compensate for the lack of visible efforts from performers \cite{schloss2003using}, so as to enrich the audience experience. The wide variety of timbres in recorded electronic music makes it difficult, if not impossible, for listeners to relate to existing musical instruments; by establishing connections between the auditory and visual domains, music visuals experienced while listening to an (electronic) music recording may help listeners to develop their internal imagery and associations with sounds.

Compared to previous works, the proposed system uses a wide range of audio features and leverages audio engineering techniques to increase creative possibilities. Whereas related systems rely on relatively simple audio representations to create visualisations, our system supports a wide range of audio features (cf. section \ref{backend}). Although this broad palette of audio features might be overwhelming for some novice users (as reflected in the user evaluation), the results discussed in the paper suggest that it can support creativity favourably for users, especially if they have some prior experience with audio processing. While feature-based music visualisation tools exist, they require explicit knowledge about coding (e.g. feature extraction and visualisation with Max/MSP and Jitter, Processing, etc.), whereas our system can be used by people with no coding experience. The insights from our study may thus be useful for future research in this domain.

The results from the user evaluation conducted for a specific song indicate that user-generated mappings yielded a significantly higher measure of complementarity between audio and visuals judged by participants, in comparison to a control condition with no mapping.
The system was found to be moderately easy to use with a mean system usability scale (SUS) \cite{brooke1996quick} measure of 58.6, indicating that certain aspects of the system may be too complex.
The system obtained a mean creativity support index (CSI) \cite{creativesupportindex} of 68.1 showing a good aptitude for exploration and expressiveness whilst the sense of immersion in the creative activity still needs to be improved. Qualitative feedback analysed through thematic analysis \cite{thematicanalysis} put forward that users liked the creative possibilities offered by the system and the synchronicity between audio and visuals in the resulting computer-generated graphics. However, issues around usability and the complexity to understand the meaning of audio features suggest that there are aspects to improve to better support users who are not familiar with MIR.

\section{Related Work}
\textbf{Music visualisation. } Torres \& Boulanger presented an agent-based framework that produces animated imagery of three-dimensional, videogame-like characters in real-time \cite{ANIMUS}. Characters in this framework were designed to respond to several types of stimuli, including sound intensity. In a subsequent study \cite{rtvisresponsiveimagery}, the authors used this framework to formulate a digital character's behaviour that responded to the emotionality of a singer's voice. Selfridge \& Barthet investigated how music-responsive visuals can be experienced in augmented/mixed reality environments for live music \cite{selfridgeaugmented}. However, their system does not enable audio feature or visual environment customisation, which is a feature of this work (see Section \ref{maindesign}).

Nanayakkara et al. presented an interactive system designed to rapidly prototype visualisations using Max/MSP and Flash \cite{experientialmusicvis}. Their visualisations were based solely on Musical Instrument Digital Interface (MIDI) signals. Although radically reducing the dimensionality of possible mappings by limiting the data types to trigger events and MIDI note numbers, this work is interesting for highlighting the use of MIDI data in the visualisation process.

Kubelka devised a hybrid system based on a combination of real-time and pre-generated visualisation techniques \cite{interactivemusicvisondrej}. It operates directly on the audio stream and maps characteristics of music to parameters of visual objects. What is notable in this study is the idea of creating a separate component for the interactive design of visualisations (a scene editor), since the majority of existing works stipulate pre-configured mappings between audio and visual properties that are not to be changed at runtime.

Music visualisation was also applied in the context of music production and learning.
McLeod and Wyvill created a software that accurately renders the pitch of an instrument or voice to a two-dimensional space, allowing a musician or singer to evaluate their performance \cite{visualizationpitch}. This example illustrates the analytic possibilities of visualisation, which become particularly apparent when viewed in educational contexts.

Systems such as Magic Music Visuals \cite{magicmusicvisuals} and Synesthesia \cite{synesthesia} represent the most recent generation of commercial audio visualisation software.
One particularly notable system is the ZGameEditor Visualizer, a toolkit integrated into the FL Studio DAW. It directly incorporates the visualisation process into the environment of the DAW.
Those commercial systems are sophisticated tools with regard to their visualisation capabilities. However, they rely on a small number of audio features, which may limit the creative potential of the visualisations.
Visual programming frameworks for music visualisation such as TouchDesigner\footnote{\url{https://derivative.ca/product}}, Max/MSP Jitter\footnote{\url{https://cycling74.com/products/max/}} or VSXu\footnote{\url{https://www.vsxu.com/about/}} do provide extensive audiovisual processing capabilities, but require significant (visual) programming skills or the use of pre-configured visualisations.

Our system differs from the previous works insofar that it can be directly integrated with a DAW. Audio engineering techniques are employed to process the incoming audio signal (cf. \ref{backend}) and provide an interactive, fully configurable mapping routine for connecting sounds to visuals. This allows for sophisticated control over the source material per se, and also the way this source material affects the visuals.

\textbf{Mapping. } \textit{Mappings} describe here virtual connections that define the behaviour of visualisations in response to audio features extracted from the music. Hunt et al. argue that the connection of input parameters to system parameters is one of the most important factors when designing digital music instruments (DMIs) \cite{parametermapping}. This consideration is worth studying for audiovisual interaction based on procedural processes. In the system presented in section \ref{maindesign}, audio features act as input parameters reacting to sound changes. When updated, they trigger a change in the system (a change in visual properties in our case). As such, the design and quality of mappings is one of the system's core considerations. 
The Open Sound Control (OSC) specification \cite{oscspecification} is one of the most popular ways of creating such mappings. 
Our system has been designed using Libmapper \cite{libmapperpaper}, an open-source, cross-platform software library based on OSC. 
The defining feature of libmapper is that mappings can be manipulated through a dedicated visual interface, while the system is actively running.

\section{Design of the System}\label{maindesign}
\subsection{Design objectives}
The production of visuals accompanying music is inherently a creative task which must take into account considerations from a number of stakeholders, e.g., artists, producers, labels. This work investigates how to design assistive tools for visual artists creating content aimed at accompanying music media. The end product is shaped by human factors (style, creative intent, a band's image and ``universe'', etc.), hence we do not target fully automatic music visualisation renderers here (e.g. the Winamp music visualisers\footnote{\url{http://www.geisswerks.com/milkdrop/}}). With this in consideration, we pose two design objectives (DO);
$DO_1$: The system should allow music producers and visual artists to create responsive visualisations based on the sonic and musical attributes from individual musical instruments, groups of instruments, or the mix as a whole. 
$DO_2$: The system should be intuitive - the process of creating mappings should be mainly visual, without extensive need of coding.

\subsection{Implementation}
\subsubsection{Backend - audio plugin}\label{backend}
The backend is a DAW audio plugin implemented in C++ using the JUCE framework \cite{storer_2020}. Its central task is audio feature extraction. Based feature extraction review by Moffat et al.~\cite{audiotoolboxoverview}, we used the \textit{Essentia} software library \cite{essentia}, as it can run in real-time and offers a large number of audio features, giving users a broad palette of possible starting points for their desired visualisations with the system. By applying the plugin to one or more tracks, an arbitrary number of backend instances can be connected to the mapping space. This provides the flexibility required by $DO_1$. In the following, we give a detailed explanation of the different aspects of the backend component.

\textbf{Feature extraction}: Feature extraction is at the core of the plugin.
Several audio features are computed globally, i.e. for whole, unfiltered chunks of incoming audio signals:

\textbf{Loudness} is an important aspect of musical expression. It provides an empirical estimate for the perceived intensity of a sound. Given that rhythmic elements can be contained in the source material, it can be employed to translate the pulse of a musical piece to periodic movements in the visualisations, for example.

\textbf{Pitch} is computed using the YIN algorithm \cite{pitchYIN}. It is one of the central features that allow listeners to distinguish between different components in music. In certain contexts it implicitly conveys aspects of emotionality \cite{pitchemotion}. This feature is most useful when the plugin is applied to isolated source material, such as an instrument or a vocal part.

The \textbf{spectral centroid}, a signal descriptor modeling the brightness perceptual attribute of a sound. It represents the frequency spectrum barycenter and has been shown to be an important correlate of timbral variations used to convey musical expression \cite{barthet2010acoustical}.

A simple \textbf{onset detection system}, based on the high-frequency content algorithm \cite{Masri1996ImrovedMO} is included. Its purpose is to recognise rapid changes in the music to identify the onset of notes or percussive instruments, which can be mapped to visual attributes. 

Lastly, the \textbf{sensory dissonance} of the signal is included in the global calculations. It measures the tension resulting from instrumental tones at a given point in time, based on the spectral peaks of the audio signal.

\textbf{Sub-bands}: In order to support $DO_1$, the system provides an option to split the incoming audio signal into up to three frequency sub-bands. These are created using second-order low- and high-pass filters respectively. Each sub-band provides a number of \textit{feature slots}, which allows for the injection of feature-computing algorithms into the given band. This is useful when the audio is a mix of several elements, e.g., a complete song. sub-bands allow the user to isolate certain parts of the signal, for example the bass on one end of the frequency spectrum and drum cymbals on the other end.

\textbf{Automatables}:
To extend the range of possible mappings, we also incorporate user-controllable metadata into the data stream. They allow users to create mappings controlling visualisations independently of the audio feature data. By utilising a feature provided by libmapper, automatables can be combined with other mappings. This gives users fine-grained control over the way a given audio signal affects a visual property. 

\begin{figure}[t]
    \begin{center}
        \includegraphics[width=0.95\columnwidth]{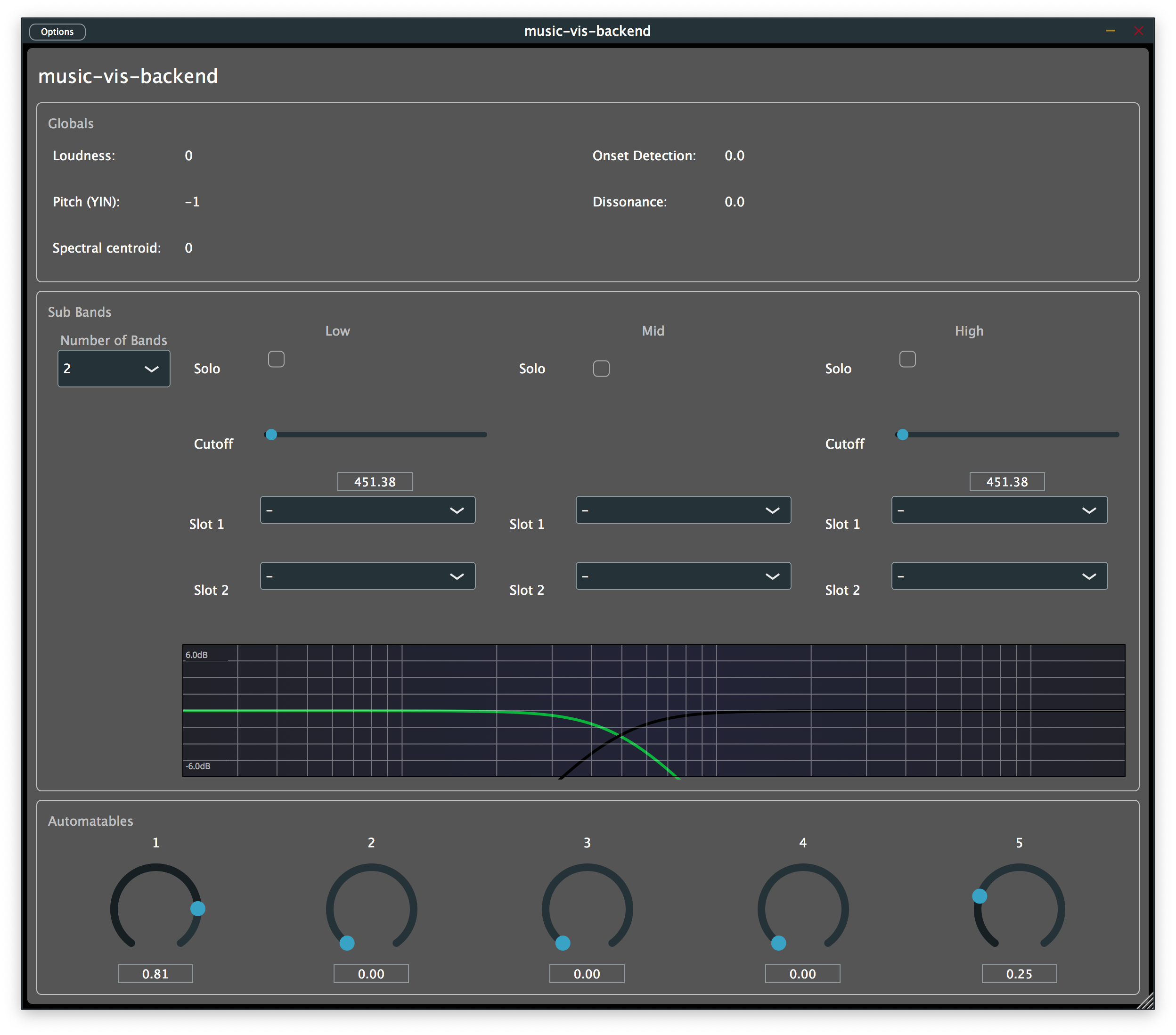}
        \caption{Screenshot of the backend GUI}
        \label{backendgui}
    \end{center}
\end{figure}

\textbf{Graphical user interface}: A GUI was created to visualise audio feature numerical data computed by the plugin and provide control over the sub-bands and automatables. 
Figure \ref{backendgui} shows a screenshot of the backend GUI.

\subsubsection{Frontend - visualisations}
The Unity\footnote{\url{https://unity.com/}} platform was selected to create the frontend due to its accessibility, extensibility as well as its potential to function on virtual and mixed reality devices.
The frontend is designed so that the visualisation engine is not bound to a specific platform. 
A software interface was created to integrate libmapper into the Unity platform. This template serves as the base class for all visual components that send or receive libmapper signals. 

\subsubsection{Middleware - libmapper}
The third cornerstone of the system is the mapping framework. Mappings between audio features and visual objects can be pre-configured, but also changed at runtime. This facilitates the creation of new roles in the context of audiovisual performances, as imagined by Bain \cite{Bain2008RealTM}.
Libmapper offers a visual interface with detailed views of signals and their connections. To support $DO_2$, we employ this browser-based GUI to make designing dataflows from the frontend to the backend a fully visual experience that does not require coding skills.
For users with basic coding skills, it offers the possibility to apply mathematical operations to signals. For the majority of audio features extracted by the backend engine, different characteristics of instruments, performers' musical expression and mix audio qualities will create differences in the numerical values output by the system. By applying operations to the data passing through the connections, signals can be tailored to an intended range or subjected to linear or non-linear transformations, for example.

\section{Evaluation}
\subsection{Procedure}
We conducted an online study to evaluate the prototype. It was structured into two tasks. Task 1 involved the observation and commentary of music videos showing visualisations produced by the system. Participants were asked to rate how well the visuals complemented the music in the videos by rating their agreement level to a 10-point Likert item (from ``not well at all'' to ``extremely well''). They were also asked to verbally describe positive and negative aspects of the visualisations. Task 1 had a suggested time limit of 15 minutes.

Task 2 of the study was an interactive user test. Participants were instructed to install the software on their own devices and completed a series of subtasks with the system, using a provided electronic music track. Subtasks included connecting certain audio features from the backend to certain visual properties of the frontend, as well as experimenting with the system on their own. For task 2, we suggested a time limit of 45 minutes. Please refer to section \ref{appsurveyquestions} for a full list of the survey questions.

\subsection{Methods}
\begin{figure}[t]
    \begin{center}
        \includegraphics[width=0.95\columnwidth]{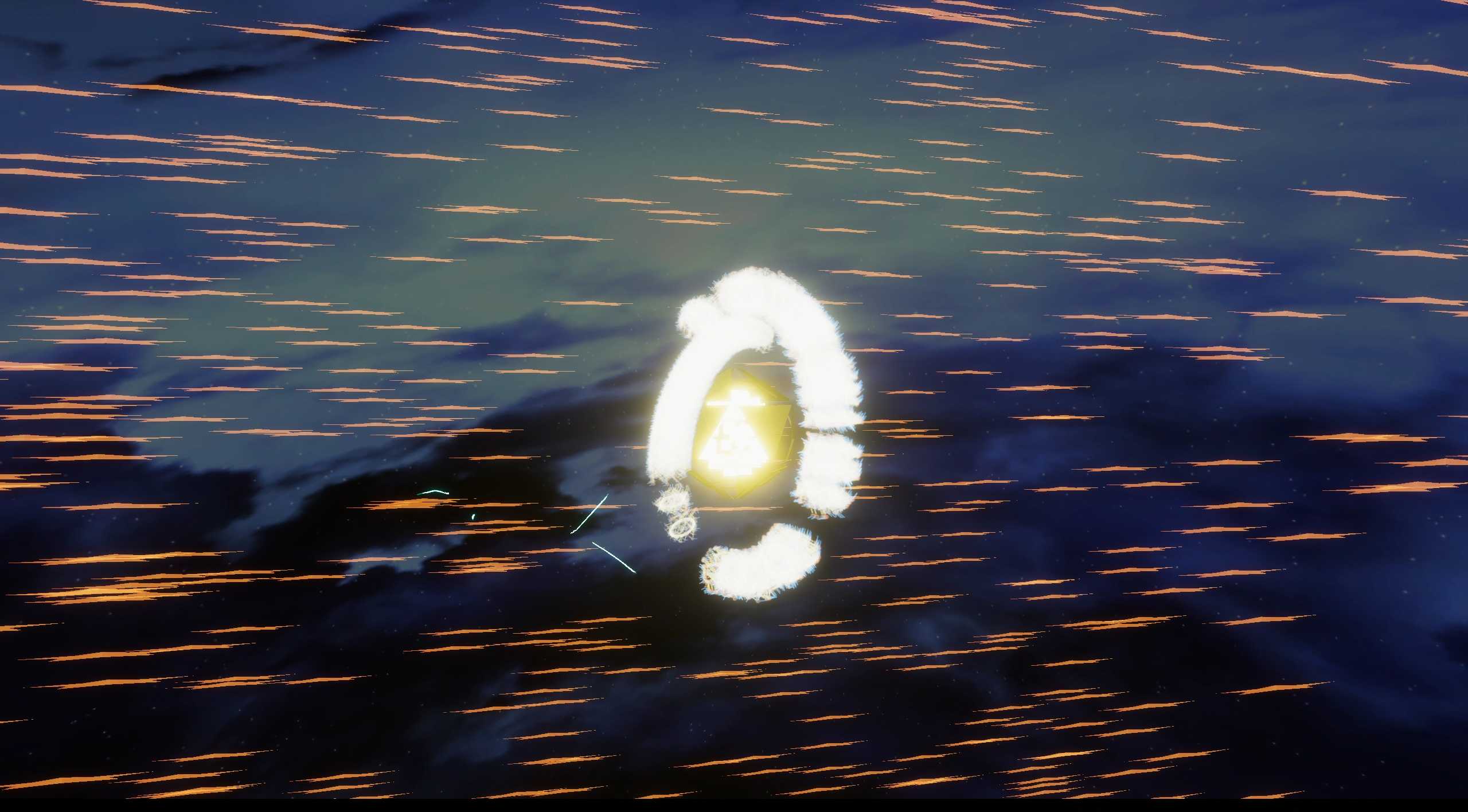}
        \caption{A screenshot of a dynamic music visulisation generated with the frontend by combining several AV mappings}
        \label{frontend}
    \end{center}
\end{figure}

To assess the visualisation capabilities of the system, we tested the following hypothesis: \textit{``Visuals based on audio-driven mappings generated by users complement the music better than visuals that do not react to the music.''}.
We designed a three-dimensional scene for the frontend that served as the foundation of mappings for the user evaluation.
Figure \ref{frontend} shows a screenshot of this scene. A link to music videos generated with the system is included in section \ref{outputvideos}.
We recorded three music videos with visualisations produced by the system for a one-minute long song. The song consists of a hip-hop style drum beat, as well as several synthesizer elements for bass and lead sounds.
Baseline automated graphic animations are incorporated into the visualisations for all three videos (e.g. slight movements of the stars). 
A music video only showing baseline animations with no audio-reactive mapping was used as control condition (M0). Two of the authors created mappings for the song independently. The mappings were added to the baseline animations provided in the control condition (M0) and yielded the two audio-driven mapping conditions, M1 and M2.
The order of the three videos was randomised across participants.

We conducted a Friedman test \cite{friedman} with the audiovisual complementarity as dependent variable and the mapping type as independent variable (three levels: M0, M1, M2).
A post-hoc analysis accounting for multiple comparisons was necessary; we used the Wilcoxon signed-rank test \cite{wilcoxon} to test differences between mapping conditions (M0-M1, M0-M2, M1-M2) by using the Bonferroni correction (the p-value significance level was \(0.017=\alpha\) based on a Type I error \(\alpha=0.05)\).

We also analysed participant feedback in task 1 by means of inductive thematic analysis based on the framework by Braun and Clarke \cite{thematicanalysis}. The goal was to identify common themes regarding positive and negative factors considering the three videos created for task 1 of the user study. 

We conducted a separate thematic analysis for the user study task 2, with the goal of finding common themes and idiosyncrasies in how users interacted with the system.
We assessed the usability of the software components using the SUS \cite{brooke1996quick}. It consists of a 10-item questionnaire, regarding topics such as complexity and ease of use.
Finally, we integrated the CSI \cite{creativesupportindex} into the study to assess the value of our system as a creativity support tool. It provides questions to measure aspects such as exploration, engagement and effort/reward trade-off.

\subsection{Participants}
34 participants took part in the study in total. No prerequisite skills were required for task 1. For task 2, we recruited participants who fulfilled certain software requirements to be able to install the software and had basic DAW skills. Participants were recruited using departmental mailing lists at our institution as well as through the \textit{Prolific} platform. We applied a custom prescreening while selecting participants, with regard to their interests (music) as well as their academic/professional backgrounds (computer science, computing, engineering or music). However, subjects were not prescreened in terms of their familiarity with music visualisation/multimedia systems specifically. Their mean age was 24.2 years (SD=5.1). 70\% were male, 30\% female, two did not disclose their gender. The majority of them were from Europe or the United States. 22 of them were students. Task 1 was completed by all participants and Task 2 by nine participants.

\section{Results}
\subsection{Quantitative Evaluation}
The results of the Friedman test showed a significant difference in the audiovisual complementarity ratings of videos based on the underlying mappings, 
\(\chi^2(2, N=34) = 15.6\), \(p < 0.001\). 

The results of the Wilcoxon signed-rank tests are listed in Table \ref{wilcoxontable}. The results indicated significant differences between M1 and M0, M2 and M0, but not between M1 and M2. Mean and interquartile range values are \(4.5\) (\(2.0-7.0\)) for M0, \(7.1\) (\(5.8-8.3\)) for M1, \(6.8\) (\(6.0-8.3\)) for M2.
This shows that visuals produced with audio-driven mappings were found to better complement the music than visuals with no audio-reactive mappings.

\begin{table}[t]
    \caption{Statistics of the Wilcoxon Test for Task 1}
    \begin{center}
        \begin{tabular}{|c|c|c|c|}
        \hline
         & \textbf{M0 - M1}& \textbf{M0 - M2}& \textbf{M1 - M2} \\
        \hline 
        Z & \(-3.563\)$^{\mathrm{a}}$ & \(-2.974\)$^{\mathrm{a}}$ & \(-0.705\)$^{\mathrm{b}}$ \\
        \hline
        p (2-tailed) & \(\mathbf{<0.001}\) & \(\mathbf{0.003}\) & \(0.481\) \\
        \hline
        \multicolumn{4}{l}{$^{\mathrm{a}}$Based on negative ranks.} \\
        \multicolumn{4}{l}{$^{\mathrm{b}}$Based on positive ranks.}
        \end{tabular}
    \label{wilcoxontable}
    \end{center}
\end{table}

\subsection{Qualitative Evaluation}
\subsubsection{Thematic Analysis}
Thematic analysis was conducted by two coders and the results were integrated. For task 1, 82 codes were extracted in total. The prevalent themes of answers obtained in task 1 were centred around the aesthetic quality of the visualisations (38 codes) and the connections between music and visualisations (28 codes). An in-depth analysis of the themes is omitted here for space reasons. For details, please refer to section \ref{appthematicanalysistask1}. The thematic analysis for user study task 2 is concerned with participants' experiences while actively using the system. The codes gathered from the answers were compiled into the following themes (code occurence numbers are reported in brackets):

\textit{\textbf{Synchronicity}} (6): 
Overall, participants were satisfied with the system's synchronicity, both in terms of latency and mappings. One participant felt that they experienced low latency when attempting one of the subtasks. Participants stated that \textit{``It followed the beat of the song well''}, \textit{``i could obtain a music visualization that fit the audio''} and \textit{``the light object in the middle of the visualisation lights up according to the music''}. One participant experienced delays between sound and visualisations (\textit{``Connection issues. its a bit laggy too''}).

\textit{\textbf{Expressiveness}} (6):
The majority could express their creative intent with the system (\textit{``The loudness and onset detection were my most favourite features since i was dealing with rhythmic music to test the system''}). The option to apply transformations to signals was well received. One participant stated: \textit{``Really liked the section to add equations to the effect response''}. Several participants described their use of mathematical operations to transform mappings (\textit{``[...] the camera movement was quite rapid but looked wonderful after diluting doing a y = 0.5*x''}, \textit{``[...] result is pretty extreme, changing the mapping function to y=0.01*x works better.''}).

\textit{\textbf{Technical Issues}} (5):
Four out of nine participants reported technical difficulties at some point during the experiment. One participant mentioned an intial unresponsiveness of the system (\textit{``The interstellar app crashed a few times upon opening but ran stably if it managed to open``}).

\textit{\textbf{Ease of Use}} (4):
Participants had a mixed experiences in terms of usability. Three participants lauded the intuitiveness of the system (\textit{``It was very intuitive and fun to use''}, \textit{``The connection was fairly simple to set up''}).
Two participants reported that they experienced difficulty in using the system because they did not understand the terminology of certain audio features. They both highlighted that there was too much “tech jargon” and that the system should be more “user friendly”. One participant claimed that they \textit{“needed more guidance on what kind of mappings would work well”}.

\subsubsection{Usability and creativity support}
The sample size for the SUS and CSI analyses was 9. Our system obtained a mean SUS score of 58.6 (SD=15.7). This is below the average score of 68\footnote{\url{https://www.usability.gov/how-to-and-tools/methods/system-usability-scale.html}}, but nevertheless shows that our system exhibits a reasonable level of usability.

\begin{figure}
    \begin{center}
        \includegraphics[width=0.95\columnwidth]{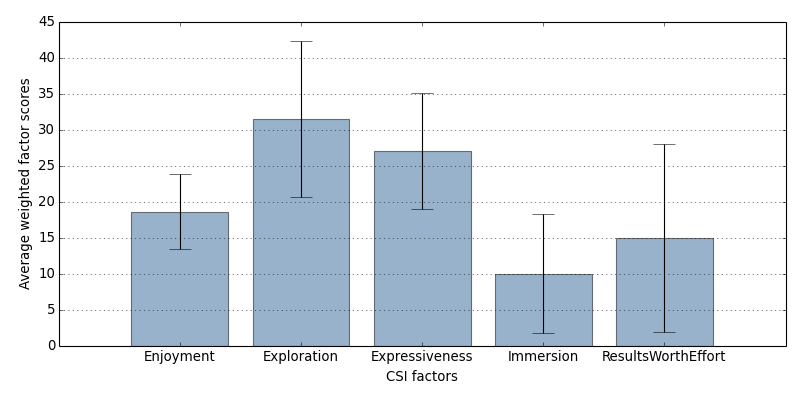}
        \caption{Bar plot of the weighted average CSI factor scores}
        \label{csifactorscores}
    \end{center}
\end{figure}

The system obtained a mean CSI score of 68.1 (SD=13.2), which indicates that its facility as a creativity support tool is above average, but not excellent. Figure \ref{csifactorscores} shows the individual weighted scores of CSI factors. The CSI results revealed that the aspects most positively received were ``Exploration'', ``Expressiveness'' and ``Enjoyment'' (in decreasing order). The factors ``Immersion'' and ``ResultsWorthEffort'' were rated rather poorly in comparison. 

\subsubsection{Duration of use in task 2}
As mentioned above, we imposed a (non-strict) time limit of 45 minutes for task 2. On average, participants worked with the system for 58.8 minutes (SD=32.6). Table \ref{durationsustable} lists the durations of use and resulting SUS and CSI scores for each participant in task 2. We subjected these data to analysis to see whether there existed an interaction between the time that participants spent working on the system and their indicated usability scores.
The working durations and SUS scores were not significantly correlated, Pearson's \textit{r}(7)=.27, \textit{p}=.47 (\textit{p}>.05, the type I error). The same was true for the working durations and CSI scores, Pearson's \textit{r}(7)=.61, \textit{p}=.07 (\textit{p}>.05, the type I error).
This indicates that there was no linear correlation between the time that users spent with the system and their given usability ratings. 

\begin{table}[t]
    \caption{Working durations in minutes (m), SUS scores and CSI scores for each participant in task 2}
    \begin{center}
        \begin{tabular}{|c|c|c|c|}
        \hline
        \textbf{Duration (m)}& \textbf{SUS score}& \textbf{CSI score}& \textbf{PID}\\
        \hline 
        31 & 30.0 & 52.3 & 9\\
        \hline
        34 & 50.0 & 80.0 & 5\\
         \hline 
        40 & 60 & 46.0 & 1\\
        \hline 
        42 & 67.5 & 76.0 & 8\\
        \hline 
        45 & 72.5 & 65.6 & 3\\
         \hline 
        50 & 62.5 & 67.0 & 2\\
        \hline 
        60 & 50.0 & 58.3 & 4\\
         \hline 
        105 & 87.5 & 81.0 & 6\\
         \hline 
        123 & 47.5 & 87.0 & 7\\
        \hline 
        \end{tabular}
    \label{durationsustable}
    \end{center}
\end{table}

\subsection{Discussion and limitations}
The results presented for task 1 of the study are limited by the fact that only one song was used. Although the results evidence a significant improvement compared to a baseline, future work should be conducted to assess whether the results generalise to other songs and other genres of music. 
It would also be worth comparing the proposed system against existing commercial and open-source software that can be used for computational music visualisation generation.

In order to improve usability, information and tutorials on audio features could be provided since most users would not be familiar with MIR, psychoacoustics and music perception (e.g., links to online support material could be provided in the plugin menu). Ways to simplify the audio-visual mapping process should be investigated e.g., by using abstractions hiding the complexity in the naming and potentially large number of audio and visual attributes, and/or interactive machine learning. 

\section{Conclusion}
We have presented a novel framework for interactively visualising sound and music. To our knowledge, this is the first system operating on the demonstrated level of interactivity. By enabling users to map every implemented audio feature to every exposed visual property, a broad range of possible visualisations is supported. We explained the system's design and implementation and discussed its application in various settings. The results of the quantitative and thematic analyses showed that music videos produced with audio-driven mappings were perceived to have a higher audio-visual complementarity than videos showing non audio-reactive visualisations. Future work could address the limitation of one song for task 1 of the user study. Analysing the audio-visual complementarity using songs from different musical genres may increase the explanatory power of the evaluation.
The results of the usability and creativity support analyses showed that there is still room for improvement to integrate help on technical audio features and to make the application less computationally expensive to reduce audiovisual lags during production.

\section{Acknowledgements}
This work has been partly funded by the UKRI EPSRC Centre for Doctoral Training in Artificial Intelligence and Music (AIM), under grant EP/S022694/1.

\bibliographystyle{jaes}
\bibliography{thesis}

\section{Appendix}
\subsection{Links to the videos created with the system} \label{outputvideos}
Videos of the three conditions created for task 1 of the user study were recorded and are available at the following link: \url{https://gofile.io/d/C1kxhI}.

\subsection{Survey Questions}\label{appsurveyquestions}
Tables \ref{questionstable} and \ref{questionstable2} list the survey questions for tasks 1 and 2, respectively, excluding the questions of the CSI and SUS analyses.

\begin{table}[t]\small
    \caption{Questions for task 1 of the survey}
    \begin{center}
        \begin{tabularx}{1.0\columnwidth}{|X|}
        \hline
        \textbf{Visualiser 1, 2 and 3 (separately for each video)}\\
        \hline
        What did you enjoy about the viewing experience? \\
        What did you dislike about the viewing experience? \\
        How well do you think the visualisation complemented the audio? \\
        What are the reasons for your answer to the question above? \\
        \hline
        \textbf{General Questions}\\
        \hline
        Would you use a system such as Interstell.AR to produce music visualisations and if so, in which context(s)? \\
        We want to improve Interstell.AR! Please report any ideas or recommendations you may have on how to improve the experience and/or what you would be interested in doing with the system. \\
        Which headphones/earphones did you use? \\
        \hline
        \textbf{Demographics and Personal Questions}\\
        \hline
        Gender?\\
        Age?\\
        Please indicate your occupation:\\
        Please indicate your nationality:\\
        Do you have any hearing impairment? If so, please specify which, if you wish.\\
        Do you have any visual impairment? If so, please specify which, if you wish.\\
        How would you describe your experience as a musician? \\
        How would you describe your experience as an audiovisual artist? \\
        \hline
        \end{tabularx}
    \label{questionstable}
    \end{center}
\end{table}

\begin{table}[t]\small
    \caption{Questions for task 2 of the survey}
    \begin{center}
        \begin{tabularx}{\columnwidth}{|X|}
        \hline 
        \textbf{Subtasks}\\
        \hline
        Subtask 1: Connect the overall loudness of the signal to the size of the particles in the centre of the scene. Please comment on the audiovisual mapping process conducted with the system and the music visualisation you obtained. \\
        Subtask 2: In this task, you will be asked to control the camera view used in the Interstell.AR visualiser based on audio onset features. Please comment on the audiovisual mapping process conducted with the system and the music visualisation you obtained.\\
        Subtask 3: Use the system to create your own mapping by experimenting with different audio features and visual elements. Please comment on the audiovisual mapping process conducted with the system and the music visualisation you obtained. Briefly describe what audio features you chose and how they manifested in the visual domain.\\
        \hline
        \textbf{General Questions}\\
        \hline
        Please describe your experience with the audio features available in the system. What additional audio features would you like to see included in the application? \\
        Do you have any other suggestions on how to further improve the system?\\
        Did you encounter any bugs or issues while working with the system?\\
        \hline
        \end{tabularx}
    \label{questionstable2}
    \end{center}
\end{table}

\subsection{Thematic analysis for task 1 of the user study}\label{appthematicanalysistask1}
For task one, the most striking positive theme regarding the two mapping conditions was the connection between music and visual elements. Approximately half of the overall answers mentioned this connection (``Consistency between audio and visuals'', ``I really enjoy the way the the visuals represent the sounds during certain sequences'', ``I enjoyed the virtual symbiosis between the effects and music [...]'').
The particle systems in the visualisation seemed to catch participants' interest the most. Six out of 34 answers stated their positive effect on the viewing experience (``The timings of the particles with the beat were perfect'', ``The floating particles in the background, i love how they react to the music'').

The main negative themes of the two mapping conditions was the lack of connections between audio and visualisations. More specifically, it appears that participants expected every visual object in the scene to be mapped to an audio feature (``Some elements weren't obviously mapping to a single feature of the audio'', ``I don't each 3d object was mapped to a single musical object [sic]'', ``[...] I didn't understand how the foggy texture related with thee [sic] music'').

Thematic analysis of the control condition led to interesting insights about the system in its default state, without any mappings applied. The most prevalent positive theme was the aesthetics of the visual scene itself (``Looks very clean'', ``The mid of the picture was nice to look at'', ``Just how HD it looks'', ``The imagery is calming''). These answers suggest that participants perceived the visual elements differently from the scenarios where they reacted to music, focusing more on the general appearances of objects.

The prevailing negative theme was the disjuncture between sound and visual objects. 21 participants commented on the stasis of the scene (``Nothing was happening in it, just repetitive'', ``Didn't feel like it reacted to the music at all'', ``It seemed way too static. No movement exept for the middle of the picture'').

\end{document}